# Generating Mixcode Popular Songs with Artificial Intelligence: Concepts, Plans, and Speculations

Abhishek Kaushik and Kayla Rush

{firstname}.{secondname}@dkit.ie

## Introduction

Music is a potent form of expression that can communicate, accentuate or even create the emotions of an individual or a collective (Juslin 2014; DeNora 2000). Both historically and in contemporary experiences, musical expression was and is commonly instrumentalized for social, political and/or economic purposes (Danaher 2010). Generative artificial intelligence provides a wealth of both opportunities and challenges with regard to music and its role in society. This paper discusses a proposed project integrating artificial intelligence and popular music, with the ultimate goal of creating a powerful tool for implementing music for social transformation, education, healthcare, and emotional well-being. Given that it is being presented at the outset of a collaboration – between a computer scientist/data analyst and an ethnomusicologist/social anthropologist – it is mainly conceptual and somewhat speculative in nature. We hope and trust that it will be received in that spirit, and we look forward to feedback from our fellow conference-goers.

Exploration of generative AI-based music is an emerging field that requires additional research. There are many reasons to investigate this location, including:

1. Investigating creative capacity: Generative AI presents a potent challenge to the traditional ethnomusicological definition of music as 'humanly organized sound' (Blacking 1973). How might generative AI redefine understandings of music, creativity, and creation? Can AI serve as a tool for musicians to innovate and explore new paths for enhanced artistic expression? Or might it foreclose innovation and displace musical creators, as many fear?

2. Investigating efficiency: Further research is required to determine how (and whether) AI-powered language models can accelerate the songwriting process, and how this can be verified across various language situations, including native or regional dialects. Given that music is culturally situated, and thus varies widely across cultural, social, and linguistic contexts, any understanding of generative AI and song generation must take into account these complexities.

3. Acceptance and inclusivity: Generative AI-based music has been heralded by some as enabling democratic music production that is accessible to all, while also providing society with personalized and customized experiences. These investigations are still not well explored. Moreover, large language models have been repeatedly shown to significantly perpetuate human biases (Navigli, Conia, & Ross 2023), and bias thus needs to be taken into account carefully when creating AI-powered language models.

4. Examining the ethical implications of merging music with artificial intelligence, including issues like copyright related to AI algorithms and the influence of AI on the music sector. Exploring these regions may provide insights on how to integrate AI and music in a more sustainable manner.

This study focuses on developing a conceptual framework for creating music based on intents. This framework centres on the widely used the mixcode language Hinglish (Yadav, Kaushik, & Sharma 2021). Hinglish is a hybrid language that merges English and Hindi features to facilitate communication in urban regions of India and other countries of South Asia. Hinglish incorporates English phrases into Hindi sentences, or vice versa (see Figure 1). It is not a standardized language; rather, it is a flexible mode of communication that lacks established guidelines (Shah & Kaushik 2019).

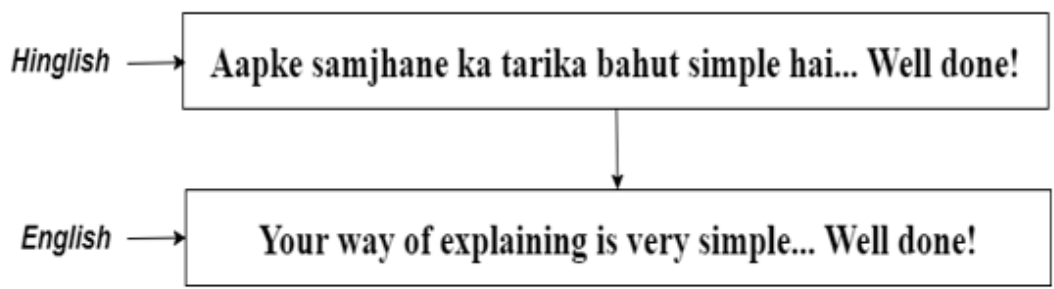

Figure 1: Example of a Hinglish statement.

Hinglish reflects South Asia's rich culture and can influence the region's large and growing population, such as the over 1.4 billion people currently living in India, the world's most populous country. Taken together, the eight member states of the South Asian Association for Regional Cooperation (SAARC) – including India, currently the world's most populous country – were estimated to have a total population of over 1.9 billion by the end of 2021 (UN Dept of Economic and Social Affairs 2022). While of course not all people in these countries use or understand Hinglish, it is worth noting that the regions in which this mixcode language is common encompass a large proportion of the global population (noting that this figure does not take into account the Hinglish-speaking global diaspora). Thus, it is vital and urgent that we as researchers pay attention to Hinglish, and its relationship (current, potential, or future) with large language models and generative artificial intelligence.

This study explores validation approaches for generative AI-based music, including human and empirical validations, using the as-yet uncharted domain of Hinglish in India. It also seeks to create a novel Hinglish dataset for training the generative model and discusses its potential commercial and social application.

Hinglish is widely spoken in South Asian countries such as India, Pakistan, Nepal, and Bangladesh for communication, education, marketing, and election campaigns. This language lacks norms, syntax, organization, records, and books, yet people nevertheless use it. When considering this framework, we decided to utilize artificial intelligence, advanced language, and generative models to establish rules or identify patterns from the vast array of data sets in comment text. The common parameters of the frameworks are categorized into three main components: 1) lyrics generator, 2) melody with an appropriate node generator, and 3) human validation. Each parameter is linked to various aspects, including the relevant data, mathematical techniques for training, and validation procedures to ensure the justification of the parameter and its integration with the data. Details are as follows (see also Figure 2):

- Generation of lyrics: The user provides their goal and purpose for the lyrics, enabling the text generation model to produce the text or lyrics accordingly. We provide a comprehensive discussion on the specifics of text production later in this paper.

- Generation of melodies: With this parameter, we will concentrate on generating a melody that complements and communicates the provided lyrics. We will merge the created song with the generated melody for human assessment. Information regarding the creation of melodies can also be found below.

- Validation processes: Evaluations can be categorized as human evaluations and other conventional machine learning evaluations. The assessment parameter is linked to both the text generation and melody generation parameters, either directly or indirectly. When generating lyrics, we can assess them using standard procedures and machine learning or human annotation. Melody generation involves standard mechanisms and human evaluation systems, both of which will be used to evaluate . The final stage of this methodology involves assessing the integration of melody production and lyrics generation with human input. This is also discussed at greater length below.

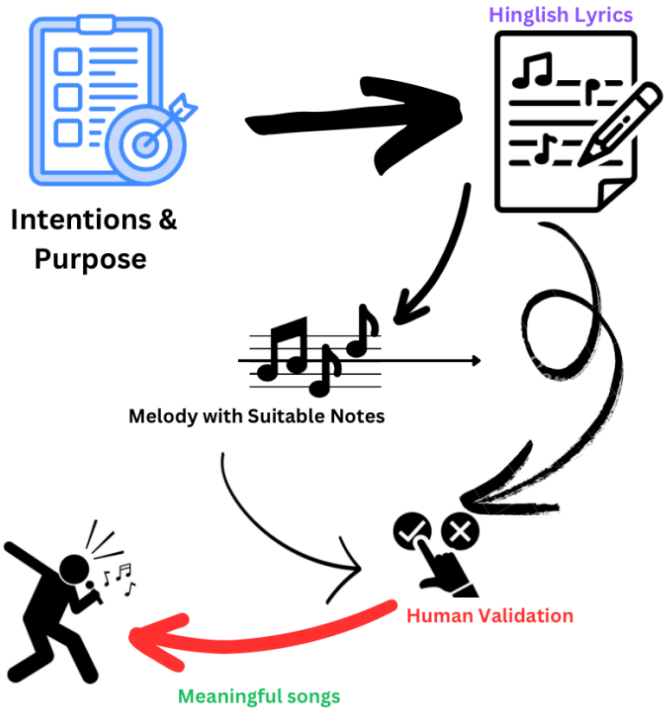

Figure 2: Flowchart of the Hinglish song generation process.

Our approach to this process is fundamentally interdisciplinary, involving two collaborators with very different backgrounds and skillsets. Creation of the generative AI for Hinglish song creation will be informed at every step by sociocultural considerations from anthropology and ethnomusicology. Chief among these is the focus on creating an AI that is culturally sensitive and appropriate for South Asian Hinglish contexts. The project is also highly experimental, charting very new ground across several areas that have been, to date, under-researched.

## Research question

Our central research question for this project is:

*How can we use artificial intelligence to generate mixcode Hinglish songs that succeed (i.e., are humanly validated) as songs within their specific cultural, social, musical, and linguistic context?*

This raises further questions that will need to be addressed along the way, such as:

- What inputs can be used to successfully train an AI in producing mixcode lyrics and melodies?

 

- Which additional aspects of songs (such as accompaniment) are absolutely necessary for a product to heard as a song, and how might these be incorporated through AI?

- How can we merge human and machine learning evaluation processes in order to judge the relative success or failure of the AI-generated Hinglish outputs?

- How might we combat the racial, cultural, linguistic, and other biases that currently plague large language models?

- What ethical and instrumental protocols must we put in place in order to use such a potentially powerful tool as a force for good in the world? (And what does that even mean?)

## Anticipated methodology and processes

As mentioned briefly above, the overall framework of the project is divided into three main components:

1. Lyrics generation. This involves the processes of natural language processing, which include data collection, data cleaning, and model training.

2. Melody generation. This component is responsible for generating melodies based on and complementary to the provided lyrics.

3. Validations. This is the stage at which we will use a combination of human validation and machine learning validation processes to judge the relative success or failure of the AI in producing mixcode Hinglish songs.

### Lyrics generation

Since this project relies on Hinglish mixcode language, we need to collect Hinglish social media text – particularly since, as discussed above, there are no formal Hinglish texts, such as books. As shown in Figure 3, after collecting the data, we must annotate it for intent classification and then train the transformer model on the annotated data for intent classification. Next, we need to gather the lyrics and high-quality data for language modelling to generate lyrics. Next, following Step 5 of the algorithm, we will combine the Indent identification and Hinglish song dataset to train a hybrid model capable of generating songs depending on the specified intent, as demonstrated in Step 6. In Step 9, we encapsulate the process of generating a song using melody generation, followed by melody validations.

> 1) Data collections to train the model for intent classifications:
> 2) Training the intent classification model with Transformer models (model ( M1 )):
>    M1 = Intent Classification Model (Transformer)
> 3) Data collection (D2 ) to generate Hinglish songs:
>    D2 = Data Collection for Hinglish Songs
> 4) Identifying the intents ( I2 ) in the data collections ( D2) with the help of (M1):
>    I2 = Identified Intents
> 5) Merging (I2) with ( D2) to form a new dataset ( D3 ):
>    D3 = Merge(I2, D2)
> 6) Training (D3) with a General Purpose Language model (GPT) resulting in (M2):
>    M2 = Training(D3, GPT)]
> 7) All the steps 2, 6 include empirical Machine learning evaluations to enhance the system: Empirical Machine Learning Evaluations
> 8) Users' intentions inputted to ( M2 ) to produce lyrics (( L)):
>    L = Input Intentions to M2
> 9) L will be embedded within the Melody generation section to produce a song (Detail discussion for melody generations and validation in later sections):
>    Song = Melody Generation(Lyrics = L)
> 10) Melody Validation

Figure 3: Algorithm for the generation of lyrics.

## Melody generation

A number of models exist for generating music through artificial intelligence. As with language models, these are rapidly improving in sophistication and complexity. With that said, the generation of music, including harmonies and accompaniment, poses several additional layers of complexity; in particular, generating music specifically for singing has not yet been well addressed (Zhu et al. 2018). In the first instance, this project focuses on the generation of melodies, as opposed to full arrangements with multiple instruments, in order to simplify the process, though accompaniment is discussed later in this section.

At minimum, an AI model for generating melodies needs to take into account pitch and rhythm, including note duration. These must be generated within a singable range, at a pace and with a rhythm that is easy to learn, ideally via aural methods since pop songs tend to be transmitted through audio and audiovisual methods, as opposed to written notation. Additionally, even where harmonies/chords are not included, popular songs tend to imply or work within the confines of more-or-less standardized chord progressions; whether these will be needed in order for a song to pass full human validation remains to be seen. We have not yet chosen our input(s) for the melody generation portion of this project, as no single choice here is ideal for our purposes; instead of presenting a process or algorithm, then, we will instead present a series of considerations that will inform our project.

On the one hand, written music notation presents the simplest, most straightforward route to machine learning in melody generation. Written notation depicts these various properties of sound such as pitch and rhythm in ways that translate easily into data or numbers. Written notation was developed in order to transfer music and sound in an era before recorded audio – thus, anyone trained in reading the notation can see the visual transcription and play or sing (or attempt to play or sing) the music as intended by the composer or transcriber. Notation softwares like Sibelius and MuseScore translates these written inputs into computer data through interfaces like MIDI, making it easy for musicians not specialized in AI and computer science (like our ethnomusicologist author) to input familiar data in ways that can be easily 'read', learned, and incorporated by the AI.

However, written notation carries with it a number of issues. The most standardized form of written notation was developed for, and continues to be most commonly associated with, Western art music (or so-called 'classical' music). Written notation is not the main medium in which popular songs are heard and shared, as these are much more likely to be learned by listening to recorded audio (Green 2001), or increasingly by watching others play and sing through video-sharing platforms like YouTube (Rush 2023). Music learning, particularly within 'intermusical' (i.e., multi- or intercultural) contexts, is a multi-sensory embodied process (van den Dool 2016), one that is difficult to replicate with AI, at least at this stage of our knowledge.

Audio input, then, may present a better way forward for this project; however, this process may be imprecise, particularly where the songs input include multiple instrumental and vocal tracks. The ideal approach for creating AI for melody generation would be to isolate sung or played melodies, but without access to a wealth of master tracks, this would present an extremely high human workload in order to ensure high-quality and accurate inputs.

While popular songs can be, and frequently are, transcribed into notated form, written notation is somewhat limited in its ability to represent all aspects of popular singing and playing. For example, syncopated melodies when transcribed often either sound clunky, losing the heard cadence or rhythm of the song, or look messy, trading in extreme degrees of specificity that are difficult to read or understand. While AIs that transcribe audio files into written notation do exist, these remain imprecise and somewhat messy, making this a less-than-ideal route for inputting data in our project. Moreover, standardized written notation is based around the Western/European twelve-tone scale; it notoriously creates difficulties when representing melodies from music-cultures with a greater number of pitch differentiations within an octave. Indian classical music is one such music-culture, and we anticipate that it will influence how Hinglish popular songs are created, heard, and performed.

Instrumental accompaniment presents an additional layer of complexity. While only a voice is needed to perform a song melody, popular songs typically feature accompaniment from one or more instruments. These instruments are somewhat standardized for pop and rock, though they do vary across music-cultures. A capella vocal melodies may not be fully validated by human listeners, given the general expectation of instrumental accompaniments for popular songs.

The final issue in melody generation is perhaps the most serious, and the one most likely to cause issues at the validation stage. A key concern is songwriting is in matching the music to the lyrics; a song understood as successful (or 'good') by human listeners tends to use melodies, harmonies, accompaniment, instrumentation, and singing style to accentuate or emphasize the lyrical and thematic meaning. Furthermore, the melodic contour and phrasing should largely match the delivery of the words and their syllables in everyday speech – and to add a further layer of complexity, pronunciation and syllabic emphasis can vary significantly by region and dialect: take, for example, the very different pronunciations of the word 'garage' in US versus European English. This careful matching of music to lyrics and/or message involves a number of different factors, including pronunciation (as discussed), but also emotion and memory.

While this section may sound a bit like a list of reasons not to attempt building an AI for melody generation for popular songs (and it admittedly felt a bit like that after writing it), we present it instead as an interesting set of challenges. A key output of this project, then, will be a research-informed algorithm for melody-for-song generation, likely informed by significant trial and error.

## Validation processes

In this project, validation is an iterative procedure subdivided into three components:

1. Validation of artificial intelligence modelling at every stage of text generation
2. Validation of generated lyrics through human annotations
3. Validation of the melody through human responses
4. Human validation of the song, which combines a created melody with lyrics

Human validation at each stage presents the most challenging aspect, as individuals hear songs differently based on a wide variety of factors, both personal and collective. We intend to perform preliminary tests of the human validation processes in a third-level songwriting module at our institution, on which we currently collaborate in teaching. (One of the paper's authors convenes this module, and the other provides guest lecturing on songwriting and AI.) As students on this module typically come from a wide variety of cultural and musical backgrounds, with varying levels of expertise and experience in songwriting and composition, this group presents a useful test group for piloting different approaches to human validation.

  

However, as our institution is located outside of the SAARC area, and this modules tends to have few or no students from South Asian backgrounds, the bulk of our human validation processes will need to take place online. We plan to engage Hinglish-speaking social media users, popular music fans, and songwriters in order to determine whether the AI-generated songs are successful as songs. Engaging with this group, moreover, will allow us to address potential ethical concerns and pitfalls, such as issues with copyright and musician concerns that AI will displace them. Particularly in light of the potential ethical implications of this project, we perceive the validation process as a mixed-methods, stakeholder-engaged approach in which we can collaborate with potential users (and skeptics) of the tool in order to chart future users that are ethical, sustainable, and culturally sensitive and relevant.

## Discussion

The convergence of artificial intelligence, data analytics, and music production has significant theoretical and practical consequences in different fields. AI and data analytics can help artists and composers create new musical ideas, melodies, harmonies, and complete works (Bryan-Kinns et al. 2024). Data analytics algorithms can analyze consumers' listening habits, tastes, and contextual information to offer personalized music recommendations. Streaming companies such as Spotify and Pandora utilize AI algorithms to recommend music based on user behaviour and interests (Kuyucu 2024). AI algorithms can assist in composing and arranging music by analyzing current musical works and creating new compositions using learned patterns and styles. OpenAI's MuseNet uses deep learning to create music across different styles and genres (Laidlow 2024). Data analytics tools can categorize music into several genres or identify stylistic aspects in songs. This skill can be utilized for activities like automatic genre labelling, music recommendation systems, and musicological study. AI systems can analyze musical elements to recognize and understand emotions transmitted in music, a process known as Emotion Recognition in Music (Kayne 2023). These findings can be applied to creating playlists depending on mood, analyzing sentiments in music reviews, and developing interactive music systems that adapt to users' emotions. The combination of AI, data analytics, and music generation offers promising prospects for improving creativity, personalization, and comprehension in music but it also brings up significant issues related to authorship, copyright, and the impact of technology on artistic expression.

Within ethnomusicology, moreover, this project raises interesting questions regarding the cultural (including regional, linguistic, and musical) specificity and contextually situated nature of songs. Developing AI that can create songs within these specific contexts – especially linguistically and musically mixcode contexts like South Asia – presents an interesting and exciting challenge, one that, if tackled sensitively, appropriately, and ethically, can inform future theorizations of music and creativity in interesting and productive ways. With that said, in addition to the challenges enumerated above, significant questions remain around combatting and neutralizing bias (especially white,

European, English-language, 'Western' bias) in AI, and these will need to be addressed explicitly within our project.

There is significant potential for impact for the AI technology created in this project. This endeavour strives to provide top-notch music spanning all genres and styles, incorporating personified music or customized music tailored to listener preferences. An opportunity exists to develop and commercialize the musical interface for musicians, producers, and aficionados to engage with the AI system. This music can be utilized in marketing initiatives aimed at musicians, producers, and content providers to spark interest and boost sales. Various monetization strategies, including one-time purchases, subscription-based services, or pay-per-use, might be considered. Music generation AI necessitates a comprehensive strategy that integrates technology advancement, user involvement, legal adherence, and ethical accountability to develop a successful and enduring product in the music sector.

In addition to the potential for commercialization, there are possible social uses for such a tool. The AI could be used to generate songs for public health and environmental campaigns and educational initiatives. The SAARC countries account for approximately 25% of the global population, and 33.4% of the global low-income population, leading to a situation in which many lack vital resources for health, wellbeing, and education (Sarkar et al. 2020). Popular mixcode songs produced by a culturally informed and sensitive generative AI could potentially be put to use by state and non-state actors in pursuit of positive societal outcomes (for example, the UN Sustainable Development Goals or similar metrics). As discussed in the beginning of this paper, music is emotionally resonant and has long been used to further social aims, and this AI tool could play a part in this. However, as with any tool, there is equal possibility of it being misused, for example by far-right actors (in any country) or by corporations and others looking to exploit vulnerable populations. This is why active engagement with Hinglish-using stakeholders is seen as a vital part of this project, as such a potentially powerful tool needs to be implemented sensitively and carefully.

We see this as a very exciting project, but even in this initial optimistic phase, we note that it presents a number of challenges, including ethical concerns, algorithmic bias, and privacy issues. Our anticipated forms of data collection are labour intensive, and the multi-stage model is complex. Focusing on a mixcode language renders the project novel and exciting, but this presents further challenges for AI in identifying patterns and creating language models based on probabilities. As discussed earlier in the paper, human validation also presents a significant challenge for this tool. In light of all of these, we see this project as a long-term investment of time and energy (and hopefully resources as well, as we are currently seeking funding from several sources). In addition to a proposed output of a powerful artificial intelligence tool, we believe this project will produce significant additions to scholarly knowledge, research, and theory, as well as important learning for future practice within this area, particularly as it relates to cultural sensitivity, bias, and ethical work. We hope to share, and look forward to sharing, future developments with all of you at AIMC conferences in 2025 and beyond.

Kashik, A., & Rush, K. (2024). Generating Mixcode Popular Songs with Artificial Intelligence: Concepts, Plans, and Speculations. AIMC 2024 (09/09 - 11/09 ). Retrieved from https://aimc2024.pubpub.org/pub/rdulfbve

# Ethics statement

This paper is primarily speculative, being presented at the beginning of a project. As such, there are no human participants involved at this stage, nor issues surrounding data management and privacy. The proposed project does present a number of potential ethical challenges, and these are discussed directly in the text of the paper. These include, for example, potential issues around copyright and ownership and ethical use of AI-generated popular songs for social purposes. We have also discussed matters of access, inclusion, and bias within the project.

Before embarking on any research, we will seek ethical approval from the relevant ethics panel at our institution.

No potential conflicts of interest were identified for this paper or this project.